\documentclass[11pt]{article}
\usepackage{graphicx}
\usepackage{amsmath}
\usepackage{amssymb}
\usepackage{amsfonts}
\usepackage{amsthm}
\usepackage{graphicx}
\begin{document}
\title{Variance matters (in stochastic dividend discount models)}
\author{Arianna Agosto\thanks{Dipartimento di Statistica Paolo Fortunati - Universit\`a di Bologna - Via delle Belle Arti 41, 40126 Bologna, Italy - e-mail: \tt{arianna.agosto2@unibo.it}} \and Enrico Moretto\thanks{(corresponding author)
Dipartimento di Economia - Universit\`a dell'Insubria - via Monte Generoso 71, 21100 Varese, Italy, Tel.: +39-0332-395510, Fax.: +39-0332-395509 - e-mail: \tt{enrico.moretto@uninsubria.it}}}
\date{}
\maketitle

\begin{abstract}
Stochastic dividend discount models (Hurley and Johnson, 1994 and 1998, Yao, 1997) present expressions for the expected value of stock prices when future dividends evolve according to some random scheme. In this paper we try to offer a more precise view on this issue proposing a closed-form formula for the variance of stock prices. \\

\end{abstract}

\section{Introduction}
\label{intro}
To determine the theoretical price of a common stock, equity valuation has always relied on the dividend discount model (DDM in the following). Such model (Williams 1938, Gordon and Shapiro 1956) determines the market value of a stock by discounting, at a suitable risk-adjusted rate, all future dividends stocks will pay to its owner. In the basic setting dividends are assumed to grow at a constant rate and the valuation formula assumes that this rate, as well as the discount one, are deterministic and constant through time.

Hurley and Johnson (1994, 1998) enhance DDM by allowing dividends to be random, following either an arithmetic or a geometric discrete-time stochastic evolution. Yao (1997) further contributes to this extension considering a trinomial evolution for future dividends and allowing the firm to go bankrupt.

More recently, Hurley (2013) imposes the growth rate of the stochastic dividend to be a continuous random variable with any given density function.

Under a theoretical point of view, this approach can be encompassed into a Markovian setting, as proved by Ghezzi and Piccardi (2003). D'Amico (2013),  applying a more general semi-Markov chain, further generalizes their results.

So far, stochastic DDM have presented expressions for the expected value of random stock prices. Needless to say, for proper investment decisions a measure of risk should be taken into proper consideration.

To offer a more precise look into stochastic dividend models, this article derives a closed-form expression for the variance of random stock prices.

The paper is organized as follows: section 2 summarizes the previous results present in existing literature while section 3 carries a closed-form expression for the variance of stock prices when future dividends follow a binomial distribution. Section 4 concludes.

\section{A binomial dividend model for stock pricing}
\label{Section 1}

In the deterministic setting DDM allows to derive a closed-form expression for the price of a stock if the following hypotheses hold:

\begin{enumerate}
\item[a)] dividends per share evolve through time at the constant geometric rate $g>-1$.
This means that the recursive equality $d_{j+1}=d_{j}\cdot \left(1+g\right)$ holds, being $d_j$ the dividends per share at time $j$. It follows that $d_j=d_0\left( 1+g\right)^{j}$, $j = 1, 2, ...$, where $d_{0}$ is the actual dividend;

\item[b)] companies are not subject to default; they will continue paying dividends forever, and

\item[c)] the discount rate $k$, i.e. the return a rational investor requires to hold the stock, is constant and such that $k>g$.
\end{enumerate}

The market value at time $0$ of the stock is, then,
$$
P_{0}=\sum_{j=1}^{+\infty }\frac{d_{j}}{\left( 1+k\right) ^{j}} =\sum_{j=1}^{+\infty }\frac{d_{0}\left( 1+g\right) ^{j}}{\left(1+k\right) ^{j}}=\frac{d_{0}\left( 1+g\right) }{k-g}
$$


Hurley and Johnson (1998) propose an interesting extension to the DDM assuming that the dividends' growth rate is a discrete random variable with $n$ possible outcomes
$$
\tilde{g}=\left\{
\begin{array}{rcccc}
\text{rates of growth} & \  & g_{1} & ... & g_{n} \\
\text{probabilities} & \  & q_{1} & ... & q_{n}%
\end{array}%
\right.
$$
where $g_{s}>-1$, $q_{s}=P\left[ \tilde{g}=g_{s}\right] >0$, $s=1,\ ...,$\ $n$, and $\sum_{s=1}^{n}q_{s}=1$. Given the current dividend per share $d_0$, future dividends at time $j$, $\tilde{d}_{j}$ are random variables themselves and follow the recursive condition $\tilde{d}_{j+1}=\tilde{d}_{j}\cdot \left( 1+\tilde{g}\right)$.

The random price of the stock is then the sum of an infinite sequence of discounted random dividends
\begin{equation}
\tilde{P}_{0}=\sum_{j=1}^{+\infty }\frac{\tilde{d}_{j}}{\left(1+k\right) ^{j}}  \label{random_P}
\end{equation}

If hypotheses b) and c) above hold, Hurley and Johnson (1998) find the following formula for the expected value of $\tilde{P}_{0}$
\begin{equation}
E\left[ \tilde{P}_{0}\right] =\frac{d_{0}\left( 1+E\left[ \tilde{g}\right] \right) }{k-E\left[ \tilde{g}\right]},  \label{HJ_Formula}
\end{equation}
as long as $k>E\left[\tilde{g}\right]$, being $E\left[\tilde{g}\right] = \sum_{s=1}^{n}g_{s}q_{s}$ the expected value of $\tilde{g}$.

It is worth noting that, in (\ref{HJ_Formula}), the expected stock price depends only on $E\left[\tilde{g}\right]$. No other moments of $\tilde{g}$ affect $E\left[\tilde{P}_0\right]$ so that, for instance, the variance of $\tilde{g}$ has no effect on $E\left[\tilde{P}_0\right]$.

\section{Variance of a stock price in the binomial case}
\label{Section 2}

The result we present in this section is two-fold: we obtain a closed-form expression for the variance of a random stock price $Var\left[\tilde{P}_0\right]$ and we show how the variance of the random rate of growth affects the variance of the stock price.

Recalling (\ref{random_P}), the variance of $\tilde{P}_{0}$ is
$$
Var\left[ \tilde{P}_{0}\right] =\sum_{j=1}^{+\infty}\sum_{p=1}^{+\infty }\frac{cov\left( \tilde{d}_{j};\tilde{d}_{p}\right) }{\left( 1+k\right) ^{j+p}}
$$
where the covariance between the stochastic dividend $\tilde{d}_{j}$ at time $j$ and the stochastic dividend $\tilde{d}_{p}$ at time $p$ comes from the standard expression
$$
cov\left( \tilde{d}_{j};\tilde{d}_{p}\right) =E\left[ \tilde{d}_{j}\cdot \tilde{d}_{p}\right] -E\left[ \tilde{d}_{j}\right] E\left[ \tilde{d}_{p} \right]
$$
Determining $E\left[ \tilde{d}_{j}\cdot \tilde{d}_{p}\right] $ requires the joint distribution of $\tilde{d}_{j}$ and $\tilde{d}_{p}$.

For sake of simplicity, in what follows we confine ourselves in the geometric case and let $\tilde{g}$ be a Bernoulli random variable
$$
\tilde{g}=\left\{
\begin{array}{ccc}
g_1 & & g_2 \\
q_1 & & q_2
\end{array}
\right.
$$
with $g_{1},g_{2}>-1$, $q_1, q_2 > 0$ and $q_1 + q_2 = 1$.

According to this choice, future dividends follow a binomial distribution
\begin{equation}
P\left[ \tilde{d}_{j}=d_{0}\left( 1+g_{1}\right) ^{s}\left( 1+g_{2}\right)^{j-s}\right] = {j \choose s} q_{1}^{s}q_{2}^{j-s}\text{,}
\label{BinProb_1}
\end{equation}
being $s=0,$\ $1,$\ $...,$\ $j$ the number of times dividend $\tilde{d}_{j}$ grows at rate $g_{1}$ in $j$ steps.

The expected value of $\tilde{d}_{j}$ is
\begin{eqnarray*}
E\left[ \tilde{d}_{j}\right] &=& d_{0}\sum_{s=0}^{j}\left(1+g_{1}\right)^{s}\left( 1+g_{2}\right)^{j-s}{j \choose s} q_{1}^{s}q_{2}^{j-s} \\
&=& d_0 \left[ \left( 1+g_{1}\right) q_{1}+\left(1+g_{2}\right) q_{2}\right] ^{j} = d_0 \left( 1+E\left[\tilde{g}\right] \right)^{j}
\end{eqnarray*}


Let also $r=0,1,...,p$ be the number of times dividend $\tilde{d}_{p}$ grows at rate $g_{1}$ in $p$ steps. It is clear that if $p\geq j$ it has to be that $r\geq s$.

The definition of conditional probability allows to find the joint probabilities of $\tilde{d}_{j}$ and $\tilde{d}_{p}$%
\begin{eqnarray*}
&&P\left[ \left( \tilde{d}_{p}=d_{0}\left( 1+g_{1}\right) ^{r}\left(1+g_{1}\right) ^{p-r}\right) \cap \left( \tilde{d}_{j}=d_{0}\left(1+g_{1}\right) ^{s}\left( 1+g_{1}\right) ^{j-s}\right) \right]  \\
&=&P\left[ \left( \tilde{d}_{p}=d_{0}\left( 1+g_{1}\right) ^{r}\left(1+g_{1}\right) ^{p-r}\right) |\left( \tilde{d}_{j}=d_{0}\left(1+g_{1}\right) ^{s}\left( 1+g_{1}\right) ^{j-s}\right) \right]  \\
&&\cdot P\left[ \tilde{d}_{j}=d_{0}\left( 1+g_{1}\right) ^{s}\left(1+g_{1}\right) ^{j-s}\right]
\end{eqnarray*}%
Such joint probabilities are strictly positive only if $s\leq r\leq s+p-j$. All other probabilities
are equal to $0$ as if in $j$ steps dividend $\tilde{d}_{j}$ has grown at rate $g_{1}$ exactly $s$ times, in the remaining $p-j$ steps dividends can grow at rate $g_{1}$ at most $p-j$ times. On the other hand dividend $\tilde{d}_{p}$ cannot have grown at rate $g_{1}$ a smaller number of time it has done already at time $j$.

It is now easy to deduce that the required conditional probabilities are
\begin{eqnarray}
&& P\left[ \left( \tilde{d}_{p}=d_{0}\left( 1+g_{1}\right) ^{r}\left(1+g_{1}\right) ^{p-r}\right) |\left( \tilde{d}_{j} = d_{0}\left(1+g_{1}\right) ^{s}\left( 1+g_{1}\right) ^{j-s}\right) \right]  \nonumber \\
&=&
\left\{
\begin{array}{ccc}
{p-j \choose r-s} q_{1}^{r-s}q_{2}^{p-j-\left( r-s\right)} & \hspace{0.2 cm} & \text{if} s\leq r\leq s+p-j \\
0                                                          &                 & \text{otherwise}
\end{array}
\right. \label{BinProb_2}
\end{eqnarray}


According to (\ref{BinProb_1}) and (\ref{BinProb_2}), the expected value $E\left[ \tilde{d}_{j}\cdot \tilde{d}_{p}\right] $ is
\begin{eqnarray*}
&&E\left[ \tilde{d}_{j}\cdot \tilde{d}_{p}\right] =\sum_{s=0}^{j}\sum_{r=s}^{s+p-j}d_{0}\left(1+g_{1}\right)^{s}\left(1+g_{2}\right)^{j-s}{j \choose s} q_{1}^{s}q_{2}^{j-s} \\
&& d_{0}\left( 1+g_{1}\right)^{r}\left(1+g_{2}\right)^{p-r} {p-j \choose r-s} q_{1}^{r-s}q_{2}^{p-j-\left( r-s\right)} \\
&=& d_{0}^{2}\sum_{s=0}^{j}\left[ \left(1+g_{1}\right)^{s}\left(1+g_{2}\right)^{j-s} {j \choose s} q_{1}^{s}q_{2}^{j-s} \right. \\
&&\left. \sum_{r=s}^{s+p-j}\left(1+g_{1}\right)^{r}\left(1+g_{2}\right)^{p-r}{p-j \choose r-s} q_{1}^{r-s}q_{2}^{p-j-\left(r-s\right)} \right]
\end{eqnarray*}%
where the inner sum is (Appendix 2)
\begin{eqnarray*}
&&\sum_{r=s}^{s+p-j}\left(1+g_{1}\right)^{r}\left(1+g_{2}\right)^{p-r} {p-j \choose r-s} q_{1}^{r-s}q_{2}^{p-j-\left( r-s\right)} \\
&=&\left( 1+g_{1}\right)^{s}\left(1+g_{2}\right)^{j-s}\left(1+E\left[\tilde{g}\right] \right)^{p-j}
\end{eqnarray*}
Now, $E\left[ \tilde{d}_{j}\cdot \tilde{d}_{p}\right] $ becomes
\begin{eqnarray*}
E\left[\tilde{d}_{j}\cdot \tilde{d}_{p}\right] &=&d_{0}^{2}\left( 1+E\left[\tilde{g}\right]\right)^{p-j}\sum_{s=0}^{j}\left( 1+g_{1}\right)
^{2s}\left( 1+g_{2}\right) ^{2\left( j-s\right) } {j \choose s} q_{1}^{s}q_{2}^{j-s} \\
&=&d_{0}^{2}\left( 1+E\left[ \tilde{g}\right] \right) ^{p-j}\left[ \left(1+g_{1}\right) ^{2}q_{1}+\left( 1+g_{2}\right) ^{2}q_{2}\right] ^{j}
\end{eqnarray*}

As
$$
\left( 1+g_{1}\right) ^{2}q_{1}+\left( 1+g_{2}\right) ^{2}q_{2}=E\left[
\left( 1+\tilde{g}\right) ^{2}\right] =1+2E\left[ \tilde{g}\right] +E\left[\tilde{g}^{2}\right]
$$
we finally get
$$
E\left[ \tilde{d}_{j}\cdot \tilde{d}_{p}\right] =d_{0}^{2}\left( 1+E\left[\tilde{g}\right] \right) ^{p-j}E^{j}\left[ \left( 1+\tilde{g}\right)^{2}\right]
$$
so that the covariance between $\tilde{d}_{j}$ and $\tilde{d}_{p}$ results being
\begin{eqnarray*}
cov\left( \tilde{d}_{j};\tilde{d}_{p}\right) &=&d_{0}^{2}\left[ \left( 1+E\left[ \tilde{g}\right] \right) ^{p-j}E^{j}\left[ \left( 1+\tilde{g}\right)^{2}\right] -\left( 1+E\left[ \tilde{g}\right] \right) ^{p+j}\right] \\
&=&d_{0}^{2}\left( 1+E\left[ \tilde{g}\right] \right) ^{p-j}\left[ E^{j}%
\left[ \left( 1+\tilde{g}\right) ^{2}\right] -\left( 1+E\left[ \tilde{g}\right] \right) ^{2j}\right]
\end{eqnarray*}

This also means that the variance of random dividend $\tilde{d}_{j}$ is
$$
Var\left( \tilde{d}_{j}\right) =d_{0}^{2}\left( E^{j}\left[ \left( 1+\tilde{g}\right) ^{2}\right] -\left( 1+E\left[ \tilde{g}\right] \right)^{2j}\right)
$$
so that eventually the covariance can be written as
$$
cov\left( \tilde{d}_{j};\tilde{d}_{p}\right) =\left( 1+E\left[ \tilde{g} \right] \right) ^{p-j}Var\left( \tilde{d}_{j}\right).
$$

Covariances between $\tilde{d}_{j}$ and $\tilde{d}_{p}$ are always positive as $1+E\left[ \tilde{g}\right] $ is positive due to the fact that $g_{1}, g_2 > -1$.

An interesting feature to stress is that dividends process, as defined here, is non-stationary. In fact the covariance between $\tilde{d}_{j}$ and $\tilde{d}_{p}$ depends not only on $p-j$ but also explicitly on $j$.

Finally, if $k>\dfrac{E\left[ \tilde{g}\right] +E\left[\tilde{g}^{2}\right] }{1+E\left[\tilde{g}\right] }$, as this condition encompasses inequality $k>E\left[\tilde{g}\right]$ (Appendix 3), the variance of $\tilde{P}_{0}$
\begin{equation}
Var\left[ \tilde{P}_{0}\right] = \dfrac{E\left[\tilde{P}_{0}\right] Var\left[\tilde{g}\right]
d_{0}\left(1+k\right)}{\left( \left(1+ E\left[\tilde{g}\right]\right) \left(k-E\left[\tilde{g}\right]\right) - Var\left[\tilde{g}\right] \right) \left(k-E\left[\tilde{g}\right]\right)}  \label{varianza}
\end{equation}
is positive and finite (Appendix 2).
%
Formula (\ref{varianza}) allows to determine how the expected stock price and the variance of the random dividends' growth rate both influence the variance of the stock price. As can be expected, the larger the variance of $\tilde{g}$, the larger the variance of $\tilde{P}_{0}$.

To conclude, we present a simple example. Assume that random future dividends of two stocks with the same current dividends per share $d_0 = 2$ and discount rate $k=0.05$ have the following random rates of growth
$$
\tilde{g}_{1}=\left\{
\begin{array}{ccc}
-0.02 & \  & 0.04 \\
0.5 & \  & 0.5
\end{array}
\right.
\hspace{0.45 cm} \text{and} \hspace{0.45 cm}
\tilde{g}_{2}=\left\{
\begin{array}{ccc}
-0.08 & \  & 0.1 \\
0.5   & \  & 0.5%
\end{array}
\right. .
$$
The expected values of $\tilde{g}_{1}$ and $\tilde{g}_{2}$ are the same: $E\left[ \tilde{g}_{1}\right] = E\left[ \tilde{g}_2\right] = 0.01$.
The expected prices of the two stocks turns out being the same as, from (\ref{HJ_Formula}),
$$
E\left[ \tilde{P}_{0}\right] =\dfrac{2\cdot 1.01}{0.05-0.01}=50.5
$$
Now, as $E\left[\tilde{g}^2_{1}\right] = (-0.02)^2 \cdot 0.5 + 0.04^2 \cdot 0.5 = 0.001$ and
$E\left[\tilde{g}^2_{2}\right] = (-0.08)^2 \cdot 0.5 + 0.1^2 \cdot 0.5 = 0.0082,$
variances of $\tilde{g}_{1}$ and $\tilde{g}_{2}$ are $Var\left[ \tilde{g}_{1}\right] = 0.001 - 0.01^2 = 0.0009$
 and $Var\left[ \tilde{g}_{2}\right] = 0.0082 - 0.01^2 = 0.0081$, being the latter is nine times greater then the first.

Formula (\ref{varianza}) allows to find variances for the two prices:
$$
Var\left[ \tilde{P}_{0}\right] =\dfrac{50.5\cdot 0.0009\cdot 2\cdot 1.05}{\left(1.01 \cdot (0.05-0.01) - 0.0009\right) \left(0.05-0.01\right)}=60.408
$$
and
$$
Var\left[ \tilde{P}_{0}\right] =\dfrac{50.5\cdot 0.0081\cdot 2\cdot 1.05}{\left(1.01\cdot(0.05-0.01)-0.0081\right) \left(0.05-0.01\right)} = 664.865
$$
that turns out being eleven times larger then the first.

\section{Conclusions}
\label{Conclusions}

In this article we have presented a closed-form expression for the variance of the price of a stock whose dividends evolve stochastically, according to a binomial scheme.

Such variance has to be intended as a companion for the expected price of the stock and these two values should be used together to determine the convenience in buying, holding, or selling a stock.

\section{Appendices}
\label{Appendices}

\subsection{Appendix 1}

The inner sum in $E\left[ \tilde{d}_{j}\cdot \tilde{d}_{p}\right] $ is
\begin{eqnarray*}
&&\sum_{r=s}^{s+p-j}\left( 1+g_{1}\right) ^{r}\left( 1+g_{2}\right)^{p-r} {p-j \choose r-s} q_{1}^{r-s}q_{2}^{p-j-\left(r-s\right)} \\
&=& \left( 1+g_{1}\right)^{s} \left(1+g_{2}\right)^{j-s} \\
&& \sum_{r=s}^{s+p-j}\left(1+g_{1}\right)^{r-s}\left(1+g_{2}\right)^{p-r-j+s} {p-j \choose r-s} q_{1}^{r-s}q_{2}^{p-j-\left( r-s\right) }\\
&=&\left(1+g_{1}\right)^{s}\left(1+g_{2}\right)^{j-s}\left[\left(1+g_{1}\right) q_{1}+\left(1+g_{2}\right) q_{2}\right]^{p-j} \\
&=& \left(1+g_{1}\right) ^{s}\left( 1+g_{2}\right) ^{j-s}\left( 1+E\left[ \tilde{g}\right] \right)^{p-j}
\end{eqnarray*}


\subsection{Appendix 2}

The variance of $\tilde{P}_{0}$ is obtained as follows
\begin{eqnarray*}
Var\left[\tilde{P}_{0}\right] &=&\sum_{j=1}^{+\infty}\sum_{p=1}^{+\infty}\dfrac{cov\left( \tilde{d}_{j};\tilde{d}_{p}\right)}{\left(1+k\right)^{j+p}} \\ &=& d_{0}^{2}\sum_{j=1}^{+\infty}\sum_{p=1}^{+\infty}\dfrac{\left(1+E\left[\tilde{g}\right]\right)^{p-j}\left[E^{j}\left[\left(1+\tilde{g}\right) ^{2}\right]-\left( 1+E\left[\tilde{g}\right] \right) ^{2j}\right] }{\left( 1+k\right)^{j+p}} \\
&=&d_{0}^{2}\sum_{j=1}^{+\infty}\dfrac{E^{j}\left[\left(1+\tilde{g}\right) ^{2}\right] -\left( 1+E\left[ \tilde{g}\right] \right)^{2j}}{\left[\left( 1+E\left[ \tilde{g}\right] \right) \left( 1+k\right) \right]^{j}}\sum_{p=1}^{+\infty }\left(\dfrac{1+E\left[ \tilde{g}\right] }{1+k}\right) ^{p} \\
&=&d_{0}^{2}\dfrac{1+E\left[\tilde{g}\right] }{k-E\left[\tilde{g}\right]}\left[\sum_{j=1}^{+\infty }\left[\dfrac{1+2E\left[\tilde{g}\right]+E\left[\tilde{g}^{2}\right]}{\left(1+E\left[\tilde{g}\right] \right)\left(1+k\right)}\right]^{j}-\sum_{j=1}^{+\infty }\left( \dfrac{%
1+E\left[ \tilde{g}\right] }{1+k}\right) ^{j}\right] \\
&=&d_{0}^{2}\dfrac{1+E\left[ \tilde{g}\right] }{k-E\left[ \tilde{g}\right]}\left(\dfrac{1+2E\left[\tilde{g}\right] +E\left[ \tilde{g}^{2}\right] }{k-E\left[\tilde{g}\right] +kE\left[ \tilde{g}\right] -E\left[ \tilde{g}^{2}\right] }-\dfrac{1+E\left[ \tilde{g}\right] }{k-E\left[ \tilde{g}\right]}\right)
\end{eqnarray*}
Summing up the two fractions into brackets results a fraction whose numerator can be written as $Var\left[ \tilde{g}\right] \left( 1+k\right)$
so that
\begin{eqnarray}
Var\left[ \tilde{P}_{0}\right] &=&\dfrac{d_{0}^{2}Var\left[ \tilde{g}\right] \left( 1+E\left[ \tilde{g}\right] \right) \left( 1+k\right) }{\left( k-E \left[ \tilde{g}\right] +kE\left[ \tilde{g}\right] -E\left[ \tilde{g}^{2} \right] \right) \left( k-E\left[ \tilde{g}\right] \right)^{2}} \nonumber \\
&=&\dfrac{E\left[\tilde{P}_{0}\right] Var\left[\tilde{g}\right]
d_{0}\left( 1+k\right) }{\left( k-E\left[ \tilde{g}\right] +k E\left[ \tilde{g}\right] - E\left[\tilde{g}^{2}\right] \right) \left( k-E\left[ \tilde{g}\right] \right)} \label{VarApp}
\end{eqnarray}
Consider the first parenthesis in the denominator of the expression above. By summing and subtracting $E^2\left[\tilde{g}\right]$ one gets
$$
k + k E\left[\tilde{g}\right] - E\left[\tilde{g}\right] - E^2\left[\tilde{g}\right] - \left( E\left[\tilde{g}^{2}\right] - E^2\left[\tilde{g}\right]\right) = \left(1 + E\left[\tilde{g}\right]\right) \left(k-E\left[ \tilde{g}\right]\right) - Var\left[\tilde{g}\right]
$$
This means that
$$
Var\left[ \tilde{P}_{0}\right] = \dfrac{E\left[\tilde{P}_{0}\right] Var\left[\tilde{g}\right]
d_{0}\left(1+k\right)}{\left( \left(1+ E\left[\tilde{g}\right]\right) \left(k-E\left[\tilde{g}\right]\right) - Var\left[\tilde{g}\right] \right) \left(k-E\left[\tilde{g}\right]\right)}
$$
\subsection{Appendix 3}

From (\ref{VarApp}) for the convergence and positiveness of $Var\left[ \tilde{P}_{0}\right]$ both
$$
k-E\left[ \tilde{g}\right] +k E\left[ \tilde{g}\right] - E\left[\tilde{g}^{2}\right]
$$
and $k-E\left[ \tilde{g}\right]$ must be strictly positive. This is the case if
$$
k>\max \left( \dfrac{E\left[ \tilde{g}\right] +E\left[ \tilde{g}^{2}\right]
}{1+E\left[ \tilde{g}\right] };E\left[ \tilde{g}\right] \right).
$$
Inequality
$$
\dfrac{E\left[ \tilde{g}\right] +E\left[ \tilde{g}^{2}\right] }{1+E\left[\tilde{g}\right] }>E\left[ \tilde{g}\right]
$$
is always true as, being equivalent to $E\left[ \tilde{g}\right] +E\left[ \tilde{g}^{2}\right] >E\left[ \tilde{g}\right] +E^{2}\left[ \tilde{g}\right] $ it boils down to $E\left[\tilde{g}^{2}\right]-E^{2}\left[\tilde{g}\right] =Var\left[ \tilde{g}\right] > 0.$

This finally means that $Var\left[ \tilde{P}_{0}\right] $ returns positive and finite values as long as
$$
k>\dfrac{E\left[ \tilde{g}\right] +E\left[ \tilde{g}^{2}\right] }{1+E\left[\tilde{g}\right]}.
$$

\end{document}